# Voyager 1 Measurements Beyond the Heliopause of Galactic Cosmic Ray Helium, Boron, Carbon, Oxygen, Magnesium, Silicon and Iron Nuclei with Energies 0.5 to >1.5 GeV/nuc

W.R. Webber[1], N. Lal[2], E.C. Stone[3], A.C. Cummings[3] and B. Heikkila[2]

1. New Mexico State University, Astronomy Department, Las Cruces, NM  88003, USA

2. NASA/Goddard Space Flight Center, Greenbelt, MD  20771, USA

3. California Institute of Technology, Space Radiation Laboratory, Pasadena, CA  91125, USA

**ABSTRACT**

We have obtained the energy spectra of cosmic ray He, B, C, O, Mg, S and Fe nuclei in the range 0.5-1.5 GeV/nuc and above using the penetrating particle mode of the High Energy Telescope, part of the Cosmic Ray Science (CRS) experiment on Voyager's 1 and 2. The data analysis procedures are the same as those used to obtain similar spectra from the identical V2 HET telescope while it was in the heliosphere between about 23 and 54 AU. The time period of analysis includes 4 years of data beyond the heliopause (HP). These new interstellar spectra are compared with various earlier experiments at the same energies at the Earth to determine the solar modulation parameter, $\phi$. These new spectra are also compared with recent measurements of the spectra of the same nuclei measured by the same telescope at low energies. It is found that the ratio of intensities at 100 MeV/nuc to those at 1.0 GeV/nuc are significantly Z dependent. Some of this Z dependence can be explained by the $Z^2$ dependence of energy loss by ionization in the 7-10 g/cm$^2$ of interstellar H and He traversed by cosmic rays of these energies in the galaxy; some by the Z dependent loss due to nuclear interactions in this same material; some by possible differences in the source spectra of these nuclei and some by the non-uniformity of the source distribution and propagation conditions. The observed features of the spectra, also including a Z dependence of the peak intensities of the various nuclei, pose interesting problems related to the propagation and source distribution of these cosmic rays.

**Introduction**

Voyager 1 (V1) has now spent 5 years in interstellar space beyond the heliosphere. The spectra of cosmic ray nuclei with Z=4-28 have already been measured in considerable detail between ~10-200 MeV/nuc (Cummings, et al., 2016). These low energy intensities may be compared with those measured at higher energies using a variety of other instruments in Earth orbit, but subject to the effects of solar modulation, in order to determine the spectral features related to the acceleration and propagation of these particles. This type of comparison has been difficult in the past because of these solar modulation effects, which are largest at low energies, and remain a problem even at energies up to ~10 GeV/nuc. In fact, the V1 measurements have shown that at ~100 MeV/nuc, the interstellar cosmic ray intensity for C and O nuclei is ~4 times that recorded by the ACE experiment in 2009 (Lave, et al., 2013) when the solar modulation level was the smallest ever recorded at the Earth.

In 2003 we reported earlier results for GeV/nuc particles from the CRS experiment on V2 (Webber, McDonald and Lukasiak, 2003). We used the particles penetrating the entire High Energy Telescope (HET) (Stone, et al., 1977) to obtain energy loss distributions of the relativistic Z=5-26 nuclei. These energy loss distributions were then deconvolved to obtain the energy spectra of the most abundant of these nuclei from ~500 to 1500 MeV/nuc along with an integral intensity value above 1.5 GeV/nuc. This earlier data was obtained in a composite time interval between 1986.0-1989.0 and 1995.0-2000.0, when V2 was between 23 and 54 AU from the Sun and at times when the solar modulation effects in the heliosphere were at a minimum. So the intensities were higher than any ever recorded at the Earth but still well below the values in the local interstellar medium (LIS). Because of the high overall statistics (~36,500 O nuclei were observed above 1.5 GeV/nuc) the deconvolution procedures could be tuned and spectra were obtained that extended the measurements made earlier at these higher energies at the Earth on HEAO-3 at a time when the solar modulation was much greater (Engelmann, et al., 1990).

In this paper we extend this higher energy technique to the almost identical V1-HET telescope using a time period of ~4 years, from 2012.9 to 2017.0, when V1 was beyond the heliopause (HP). The energy range covered by these measurements, -0.5-1.5 GeV/nuc, is between the lower energy V1 measurements recently published (Cummings, et al., 2016) and measurements at higher energies where solar modulation becomes unimportant (~30-100 GeV/nuc) and where high precision intensities are becoming available from the AMS-2 experiment (e.g., AMS at CERN, 2015).

## The Data

Figure 1 shows a drawing of the bi-directional HET telescopes that are on V1 and V2. We are interested here in the particles that enter the bottom half of the telescopes through the counters B1 and B2. These particles then pass through C2, C3, C4≡C2-4 and C1 and a PEN event is registered as B1, C2-4 and C1, a 3 dimensional analysis of the three recorded pulse heights B1, C1 and C2-4. C2-4 represents the sum of C2+C3+C4, each counter consisting of a pair of 3mm thick counters (see Figure 1). In a cross plot matrix, B1 vs. C1 or C2-4 vs. C1 pulse height events, both pulse heights should lie along the diagonal of the corresponding pulse height matrix as is illustrated in Figure 2. Thus consistency criteria between B1 and C1 can be applied, for example, resulting in a C2-4 vs. C1 matrix in which the individual charges lie along a diagonal which is determined by their energy loss≡energy in each detector. This is the region between the two heavy diagonal lines in Figure 2. This energy loss is determined to a high precision in the total of 1.8 cm of Si detectors forming C2-4. This results in pulse height distributions such as those in Figures 3, 4 and 5 of Webber, McDonald and Lukasiak, 2003, (W, M and L here after) showing the individual charges. The FWHM for relativistic O nuclei is ~7.5%. In this paper, in Figure 3, we show corresponding results for B through O nuclei from the 4 year time period when V1 was beyond the HP.

The analysis of the individual energy spectra for each charge, is best illustrated by the O distribution of events in Figure 3 which extends from the highest O energies which result in the lowest channels, up through the maximum in the pulse height distribution, $I_{max}$, up to pulse heights ~1.5 $I_{max}$. The deconvolution procedure (pulse height vs. energy) of this pulse height distribution is described in detail in appendix A of W, M and L. The W, M and L spectrum was unfolded out to pulse heights ~1.50 times the minimum O pulse height, at which point it overlaps the distributions for relativistic Ne nuclei. This corresponds to a low energy ~450 MeV/nuc for O nuclei.

This unfolding of the pulse height distribution for each charge used in this paper is similar to the unfolding of the Cerenkov pulse height distributions used by Engelmann, et al., 1990 (see e.g., Figure 4 of that paper), to obtain the energy spectra of Be-Ni nuclei from ~0.6 to ~12 GeV/nuc in their seminal HEAO-3 experiment. In the Voyager experiment we are dealing with the energy loss distributions in 1.8 cm of Si, which are governed by Landau distributions whereas in the HEAO-3 experiment the variations are mainly Gaussian distributions governed by the number of Cerenkov photo electrons. In our experiment the fact that the ionization energy loss for each charge passes through a minimum at between 2.5 and 5 GeV/nuc and then flattens out at higher energies up to ~10 GeV/nuc and beyond (see Figure 13 of W, M and L), sets an upper limit at which the energy loss is changing systematically as a function of energy, of about 1.5 GeV/nuc, above which only an integral flux may be obtained. In the HEAO-3 experiment, because of the dependence of the number of Cerenkov photo electrons on energy, this limit is about ~12 GeV/nuc with an integral point at higher energies.

We also have another limitation which prevents us from obtaining a full energy spectrum from 450 MeV/nuc to >1.5 GeV/nuc for all charges. This is a background for a specific charge Z which arises mainly from the (Z-1) charge (see Figures 3, 4 and 5 of W, M and L). These (Z-1) particles lie between the two heavy lines in Figure 2 and involve (Z-1) particle of energies ~200-400 MeV/nuc and below which have a higher relative energy loss than charge Z particles at higher energies. At still lower energies these lower Z particles will eventually "pull off", above or below the diagonal criteria that are set, depending on whether they are "forward" or "backward" moving particles and as a result they are no longer a background for the higher Z particles. In fact, these particles may be analyzed separately to determine the lower energy spectra of some of the forward moving charges. A detailed study of this lower Z "background", which depends mainly on the spectra of Z-1 nuclei below a few hundred MeV/nuc, was made in W, M and L. Here we select charges where this estimated background from (Z-1) particles is less than 15% of the total counts per channel of the Z nuclei at the minimum ionizing peak, thus including only B, C, O, Mg, Si and Fe nuclei in this analysis. In each case the Z-1 charge has an abundance less than ~25% of the charge for which the spectrum is being obtained and the background from these lower energy events is ~10-15% or less and can be estimated from the spectra of the Z-1 nuclei.

The C2-4 pulse height distributions, an example which is shown in Figure 3 in this paper, may be used to obtain a peak pulse height channel for each of the above six nuclei. These peak channels are presented in Table 1. Also included in the table are the peak channels obtained from the W, M and L analysis for the V2 telescope. It is seen that the pulse height ratio between the V1 and V2 telescopes in the C4 counter is, V1/V2=0.909+0.005, an average accuracy of less than ~1%. So to derive the energy spectra on V1 we can use the W, M and L energy vs. energy loss calculations (e.g., pulse height channel #'s) for V2, x 0.909=V1 for all six charges. The measured peak E-loss channel for each charge occurs at a pulse height equal to ~1.02 times the channel which defines the 1.5 GeV/nuc energy loss, so this peak channel that is determined from the energy loss distributions essentially defines the energy loss vs. energy scale for each charge.

Tables 2A and 2B show the energy interval, the corresponding pulse height interval, the counts and various correction factors, including the deconvolution factor and the background subtraction, leading to the intensity values, percent errors, and various charge ratios derived from the V1 data in this paper for B, C, O, Mg, Si and Fe nuclei.

Table 3 in this paper shows the intensities of the nuclei presented in Tables 2A and 2B as a fraction of the Oxygen nuclei intensity=100, corresponding to Table 4 in W, M and L and also Table 4 in Engelmann, et al., 1990.

In W, M and L, Appendix B, it was demonstrated that a self-consistent determination of an average spectral index, $\gamma$, for each charge could be obtained at an energy of 1.5 GeV/nuc from the relation $\gamma=(1+1.5j_0/J_0)$, where $j_0$ is the measured differential intensity at 1.5 GeV/nuc and $J_0=$ the measured integral intensity above 1.5 GeV/nuc. These calculations of the values of $\gamma$ are shown individually in Table 4 for B, C, O, Mg, Si and Fe nuclei in the present case for V1. The new values of $\gamma$ are the same within $\pm5\%$ of those derived by W, M and L.

The intensity values at all energies for these six nuclei that we obtain from this new V1 analysis as listed in Tables 2A and 2B are shown in Figures 4, 5, 6, 7, 8 and 9. These figures include the LIS cosmic ray intensities at lower energies for these six nuclei reported by Cummings, et al., 2016, and also other spectra measured inside the heliosphere above ~0.5 GeV/nuc including the earlier W, M and L and HEAO-3 results. The spectra between 1.5 and 2.8 GeV/nuc that are obtained from the spectral index, $\gamma$, in the present paper are shown as solid red lines between 1.5 and 2.8 GeV/nuc. The 2.8 GeV/nuc point is the energy at which, on average, 0.50 of the integral counts >1.5 GeV/nuc are expected for the differential spectral index obtained which is ~2.0 or slightly less.

The intensities obtained by Engelmann, et al., 1990, for all six nuclei, and also from PAMELA for B and C nuclei (Adriani, et al., 2013) are included in the above Figures. These earlier measurements at the Earth illustrate the considerable level of solar modulation which ranges, for the HEAO-3 data for O nuclei, from a factor of nearly 3.0 at 620 MeV/nuc to a factor ~1.5 at 1.5 GeV/nuc.

The derivation of the He nuclei spectrum represents a special case of the spectral analysis for the heavier nuclei described above. In this case the spectrum is unfolded from the pulse height distribution along the C2-4 vs. C1 diagonal as is the case for the heavier nuclei. The He counts are obtained in the high gain mode where the peak channel of the pulse height distribution is now 29.5. This pulse height distribution is shown in Figure 10 along with the energy scale that is established from the value of the peak channel and also from the channel value of 207 which corresponds to the maximum stopping energy of 74 MeV/nuc. These channel and energy values are shown in Table 5.

The complete He spectrum may be determined out to energy losses ~3.2 times minimum from the distribution of diagonal events shown in Figure 10. The number of events in each energy loss interval, along with the deconvolution correction factor, interaction correction factor and geometry factor x time correction are used to obtain the final intensity. At energy losses >3.2 times minimum, which is about 140 MeV/nuc, the forward and backward He events begin to move off of the diagonal criteria set on the C2-4 vs. C1 matrix. This part of the He pulse height distribution is shown in Figure 11. Superimposed on this distribution is the energy scale which matches the stopping channel of 207 for 74 MeV/nuc. Only in the energy region between ~130-140 MeV/nuc where the forward (F) and backward (B) moving particles separate from the diagonal, is the spectrum undetermined.

So, in total, the He spectrum can be obtained continuously from about 80 MeV/nuc to ~1500 MeV/nuc.  This spectrum is shown in Figure 12.  It agrees with the previous He spectrum from ~130 to 550 MeV/nuc presented by Cummings, et al., 2016, to within ~5-10% but extends this spectrum to higher energies.

The most difficult aspect of the determination of the He spectrum is the background from the forward and backward moving (Z-1) particles, that is, protons.  This involves protons of energy losses from ~3.2-4.5 times minimum, corresponding to proton energies of 100-140 MeV.  The flux of these protons is ~60 p/m$^2$·sr·s, comparable to the flux of relativistic He above 1500 MeV/nuc.  But these particles are well separated using the B1 vs C1 and C2-4 vs. C1 criteria that we establish (see Table 5).

## The Solar Modulation Potential, ϕ, That is Present in Earlier Earth Based Experiments as Inferred from the New Voyager GeV/nuc Measurements

The average ratios of intensities at 700 MeV/nuc measured at V1 beyond the heliopause to those at the Earth by HEAO-3 for C and O nuclei are 2.90 $\pm$ 0.10 and 3.0 $\pm$ 0.10, respectively.  Calculations of the solar modulation required to produce a ratio of 2.95 using a purely "force field" modulation model (e.g., Usoskin, et al., 2011) give a modulation potential ϕ=660 $\pm$ 40 MV at the time of the HEAO-3 measurement.   In their original paper, Engelmann, et al., 1990, estimate an average modulation potential of 600 MV for extrapolation to the interstellar spectrum.

In Figure 6, which shows the newly measured O nuclei spectrum, we also show as a dashed line the predicted HEAO-3 spectrum for a modulation potential = 660 MV starting with an LIS spectrum similar to ours.  The agreement with the HEAO-3 measurements is excellent.

Also in Figure 5, which shows the newly measured C nuclei spectra beyond the heliopause, we show as dashed lines the predicted spectra at the Earth for modulation potentials of 400 MV and 660 MV which fit the PAMELA and HEAO-3 data respectively.

## Calculation of the Spectra in a LBM for Cosmic Ray Propagation in the Galaxy and a Comparison of Voyager Data at ~100 MeV and 1.0 GeV/nuc

The LBM in its simplest form provides a remarkably accurate description of the observed spectra of galactic cosmic rays (e.g., Schlickeiser and Lerche, 1985). It is therefore valuable to compare the LBM calculations with the new local spectra of cosmic ray nuclei measured by Voyager at lower energies (Cummings, et al., 2016), and those reported in this paper which are in the energy range between ~0.5-1.5 GeV/nuc. Deviations between the predictions and measurements may be used to examine departures from the LBM assumptions, therefore providing information on the origin, acceleration and propagation of cosmic rays in the galaxy.

For calculations of these spectra and the relative intensities between ~100 MeV/nuc and 1.0 GeV/nuc we use a truncated LBM, with a path length $\lambda = \rho\beta T = 18.2 \ \beta \ P^{-0.45}$ where $\rho$ is the density and T = the diffusion life-time which is $\sim K^{-1}$. Two of the most important parameters in this calculation are the source spectral indices of the spectra of the various nuclei and the value and the rigidity dependence of the diffusion coefficient used in the propagation in the galaxy. For the source spectra we assume that all of the primary nuclei C, O, Mg, Si and Fe, have the same source rigidity spectrum with a spectral index, $\gamma$, which is independent of rigidity and equal to between -2.24 and -2.28, e.g., $dj/dP = P^{-2.26}$, throughout the rigidity range.

For the diffusion coefficient K, we assume that K $\sim P^{0.45}$ above about 0.5 GV (>33 MeV/nuc for nuclei with A/Z=2). This particular combination of source and propagation spectral indices gives spectra at high rigidities, e.g. above ~100 GV, which are $\sim P^{-(\gamma+0.45)}$, where $\gamma$ is the source spectral index. As a result, for a source spectral index = -2.26, interstellar spectra that are $\sim P^{-2.71}$ are to be expected in this higher rigidity range, which is indeed what is observed for He (AMS at CERN, 2015). The dependence of K(P) $\sim P^{0.45}$ is also consistent with the observed rigidity dependence of the B/C ratio measured by AMS-2 between ~0.5 and 1000 GV (see e.g., Webber and Villa, 2016, 2017).

A casual inspection of the measured spectra in Figures 4-9 shows many similarities between the spectra. However, if they are all normalized at, for example, the energy of 1 GeV/nuc, then a consistent pattern of spectral difference emerges at lower energies. This pattern is shown in Figure 13. The particles of higher Z nuclei are deficient in low energy particles relative to those of lower Z.

In Figure 14 we have compared the calculated and measured intensity ratios between 100/1000 MeV/nuc for He, C, O, Mg, Si and Fe nuclei. The calculations, shown as black horizontal lines, are the calculated ratios for source rigidity spectra with indices = -2.24, -2.26 and -2.28 for source spectral indices that are independent of rigidity. The calculations are made with a single break in the rigidity dependence of K(P) at $P_0$=0.562 GV. Below this rigidity, K(P) ~$P^{-1.0}$. The calculations are made for a uniform LBM where the PLD distribution is an exponential as described by Schlickeiser and Lerche, 1985, and also for a modified LBM in which the PLD distributions are truncated at small path lengths (e.g., Garcia-Munoz, et al., 1987). This effect could arise, for example, if the source distribution were non-uniform or the propagation conditions were non-uniform or a combination of both. This truncation may be approximated by a double leaky box model (DLBM) in which the propagation is the sum of two exponentials, $\lambda_1$ and $\lambda_2$, as described by Cowsik and Wilson, 1975. The resultant PLD is

$$P(x)=[e^{-x/\lambda.1}+ e^{-x/\lambda.2}]$$

We have taken $\lambda_2$=0.04 for He and 0.12 for C and heavier nuclei. These different truncation parameters are in proportion to the rates of energy loss by ionization ($Z^2$/A) for He and C. These parameters provide the calculated 100/1000 MeV/nuc ratios shown in Figure 14 as well as the intensities shown in Figures 4-9 and Figure 12 for He, C, O, Mg, Si and Fe.

The measurements of the ratios, shown as red horizontal lines for each charge with error bars, generally lie near the low end of the range of the ratios that are calculated using the DLBM for the source spectral indices in the ranges -2.24 to -2.28, e.g., they are most consistent with source spectra with an exponent = -2.24. The differences in the ratios for different nuclei are due primarily to several factors; (1) the $Z^2$/A dependence of energy loss by ionization in the interstellar medium; (2) the conversion of rigidity source spectra to measured energy/nuc spectra e.g., Cummings, et al., 2017; (3) the interaction correction in interstellar H and He for the different path lengths at the two energies 100 and 1000 MeV/nuc; and (4) truncation effects due to the details of the source distribution.

At still lower energies below 100 MeV/nuc, the intensity differences between lower and higher Z nuclei increases significantly. The $Z^2/A$ dependence of energy loss by ionization is even more important at those lower energies, but other charge dependent effects may be important as well. This lower energy range has been discussed earlier in terms of the truncation of the assumed exponential PLD at short path lengths (Webber and Higbie, 2015) and will be treated in a separate paper including the He/C ratio and the individual He and C spectra from a few MeV/nuc to ~1 TeV/nuc (Webber, 2017).

**Summary and Conclusions**

This paper describes a determination of the energy spectra of galactic cosmic ray nuclei at energies between about 0.5-1.5 GeV/nuc of several of the most abundant cosmic ray nuclei measured at V1 beyond the heliopause. These energy spectra are determined from the energy loss distributions in a 1.8 cm thick array of three Si detector pairs using a similar procedure to that used to determine these spectra at V2 when it was inside the heliosphere (W, M and L). The energy range over which the differential spectra are measured is from ~500-1500 MeV/nuc with an integral flux obtained above ~1500 MeV/nuc.

These are the first measurements of the LIS spectra in the GeV/nuc energy range which are free of solar modulation effects. As such they may be compared with measurements made near the Earth to establish improved values for the modulation parameters appropriate to these Earth based measurements.

These new GeV/nuc measurements may also be compared with lower energy measurements at ~100 MeV/nuc that are made on Voyager (Cummings, et al., 2016). When this is done it is found that the measured intensity ratios between 100 and 1000 MeV/nuc are consistent with propagation in a truncated LBM with source spectra that are ~$P^{-2.24\pm 0.02}$, and are the same for each of the nuclei and also with this source exponent independent of rigidity over the corresponding rigidity range. In this DLBM the diffusion coefficient is taken to be ~$P^{0.45}$ above 0.562 GV.

The charge dependence of this ratio that is shown in Figure 13 above 100 MeV/nuc is caused mainly by the $Z^2/A$ dependence of ionization loss in the interstellar medium. The values of this ratio are also dependent on the spectral index of the source spectra and the conversion to energy/nuc spectra that are measured. The effects can be seen in both Figures 13 and 14.

These measured spectral differences below 100 MeV/nuc correspond to systematic energy dependent peaks in the differential energy spectra of the various nuclei at energies <100 MeV/nuc, perhaps not completely explained by the direct $Z^2/A$ ionization loss effects but by differences in the path length distributions (truncation) and other Z dependent effects for each charge. These differences will be considered in separate papers concentrating on the lower energy particles, both primary and secondary.

**Acknowledgements:** The authors are grateful to the Voyager team that designed and built the CRS experiment with the hope that one day it would measure the galactic spectra of nuclei and electrons. This includes the present team with Ed Stone as PI, Alan Cummings, Nand Lal and Bryant Heikkila, and to others who are no longer members of the team, F.B. McDonald and R.E. Vogt. Their prescience will not be forgotten. This work has been supported by NASA Grant NNN12AAO1C.

| TABLE 1 | | | |
|---|---|---|---|
| **Observed Peak Channel of the Pulse Height Distributions in C2-4** | | | |
| **Charge** | **W, M, & L** | **This Paper** | **Ratio** |
| B | 42.2 | 38.3 | $0.908 \pm 0.017$ |
| C | 60.6 | 55.0 | $0.908 \pm 0.004$ |
| O | 108.5 | 98.6 | $0.909 \pm 0.004$ |
| Mg | 246.0 | 223.7 | $0.909 \pm 0.006$ |
| Si | 338.5 | 308.0 | $0.910 \pm 0.006$ |
| Fe | 1176 | 1071 | $0.911 \pm 0.007$ |

**TABLE 2A**
**B PEN DIAGONAL**
**BORON**

| ENERGY | Min=38.3 | Channel # | Cnts | Back | Cnts/MeV | Decon | Interact | (x GF x t)$^{-1}$ | +/- % | B/C |
|---|---|---|---|---|---|---|---|---|---|---|
| >1500 | 1.050 | 40.25 | 605 | 59 | 546* | 1.22 | 1.09 | 1.110* | 5 | 0.307 |
| 1200-1500 | 1.092 | 41.80 | 240 | 19 | 0.737 | 0.67 | 1.09 | 0.820 | 8 | 0.292 |
| 1000-1200 | 1.136 | 43.48 | 188 | 13 | 0.875 | 0.79 | 1.09 | 1.146 | 9 | 0.301 |
| 800-1000 | 1.205 | 46.00 | 225 | 15 | 1.05 | 0.91 | 1.09 | 1.60 | 8 | 0.289 |
| 600-800 | 1.312 | 50.11 | 276 | 16 | 1.300 | 0.96 | 1.09 | 2.10 | 7 | 0.267 |

**CARBON**

| ENERGY | Min=55.0 | Channel # | Cnts | Back | Cnts/MeV | Decon | Interact | (x GF x t)$^{-1}$ | +/- % | C/O |
|---|---|---|---|---|---|---|---|---|---|---|
| >1500 | 1.046 | 57.53 | 2032 | 220 | 1812* | 1.18 | 1.11 | 3.61* | 3 | 1.043 |
| 1200-1500 | 1.082 | 59.51 | 796 | 89 | 2.36 | 0.70 | 1.11 | 2.80 | 5 | 1.048 |
| 1000-1200 | 1.121 | 61.66 | 618 | 65 | 2.76 | 0.815 | 1.11 | 3.81 | 6 | 1.024 |
| 800-1000 | 1.190 | 65.43 | 792 | 77 | 3.57 | 0.920 | 1.11 | 5.56 | 5 | 1.047 |
| 600-800 | 1.307 | 71.88 | 1061 | 100 | 4.81 | 0.97 | 1.11 | 7.88 | 4 | 1.072 |

**OXYGEN**

| ENERGY | Min=98.6 | Channel # | Cnts | Back | Cnts/MeV | Decon | Interact | (x GF x t)$^{-1}$ | +/- % | O/Si |
|---|---|---|---|---|---|---|---|---|---|---|
| >1500 | 1.043 | 102.84 | 2306 | 576 | 1730* | 1.15 | 1.125 | 3.46* | 4 | 5.35 |
| 1200-1500 | 1.076 | 106.10 | 861 | 212 | 2.17 | 0.72 | 1.125 | 2.67 | 6 | 6.30 |
| 1000-1200 | 1.114 | 109.84 | 692 | 169 | 2.62 | 0.83 | 1.125 | 3.72 | 7 | 5.76 |
| 800-1000 | 1.178 | 116.15 | 866 | 196 | 3.35 | 0.925 | 1.125 | 5.31 | 6 | 5.61 |
| 600-800 | 1.301 | 128.28 | 1082 | 205 | 4.40 | 0.976 | 1.125 | 7.35 | 5 | --- |
| 500-600 | 1.410 | 139.03 | 668 | 130 | 5.34 | 0.990 | 1.125 | 9.56 | 6 | --- |
| 440-500 | 1.500 | 148.00 | 483 | 83 | 6.66 | 1.00 | 1.125 | 11.4 | 1 | --- |

Criteria are: B1 > 16; C1 > 12; $(-2 + 1.48 \times C1) \leq B1 \leq (2 + 1.89 \times C1)$

| TABLE 2B MAGNESIUM | | | | | | | | | | |
|---|---|---|---|---|---|---|---|---|---|---|
| ENERGY | Min=223.7 | Channel # | Cnts | Back | Cnts/MeV | Decon | Interact | (x GF x t)$^{-1}$ | +/- % | Mg/Si |
| >1500 | 1.038 | 232.5 | 466 | 76 | 390* | 1.13 | 1.14 | 0.765* | 6 | 1.205 |
| 1200-1500 | 1.071 | 240.0 | 171 | 27 | 0.480 | 0.77 | 1.14 | 0.632 | 12 | 1.49 |
| 1000-1200 | 1.108 | 248.1 | 144 | 23 | 0.605 | 0.88 | 1.14 | 0.921 | 12 | 1.43 |
| 800-1000 | 1.175 | 263.2 | 187 | 24+(32) | 0.650 | 0.95 | 1.14 | 1.074 | 10 | 1.15 |

| SILICON | | | | | | | | | | |
|---|---|---|---|---|---|---|---|---|---|---|
| ENERGY | Min=309.0 | Channel # | Cnts | Back | Cnts/MeV | Decon | Interact | (x GF x t)$^{-1}$ | +/- % | Si/Fe |
| >1500 | 1.035 | 320.0 | 436 | 100 | 336* | 1.10 | 1.15 | 0.647* | 9 | 1.87 |
| 1200-1500 | 1.066 | 330.1 | 120 | 28 | 0.307 | 0.79 | 1.15 | 0.424 | 13 | 1.61 |
| 1000-1200 | 1.104 | 342.4 | 108 | 26 | 0.410 | 0.90 | 1.15 | 0.646 | 13 | 1.66 |
| 800-1000 | 1.173 | 363.7 | 160 | 32+(20) | 0.540 | 0.965 | 1.15 | 0.912 | 13 | 1.50 |
| 640-800 | 1.270 | 392.4 | 130 | 24 | 0.663 | 0.995 | 1.15 | 1.155 | 13 | 1.58 |

| IRON | | | | | | | | | | |
|---|---|---|---|---|---|---|---|---|---|---|
| ENERGY | Min=1076 | Channel # | Cnts | Back | Cnts/MeV | Decon | Interact | (x GF x t)$^{-1}$ | +/- % | O/Fe |
| >1500 | 1.032 | 1110 | 195 | 15 | 180* | 1.07 | 1.18 | 0.346* | 10 | 10.0 |
| 1200-1500 | 1.061 | 1141 | 54 | 5 | 0.163 | 0.89 | 1.18 | 0.264 | 18 | 9.92 |
| 1000-1200 | 1.098 | 1181 | 50 | 5 | 0.225 | 0.96 | 1.18 | 0.388 | 18 | 9.60 |
| 800-1000 | 1.177 | 1266 | 81 | 7+(5) | 0.345 | 0.985 | 1.18 | 0.610 | 16 | 8.71 |
| 600-800 | 1.317 | 1418 | 92 | 7+(5) | 0.400 | 1.00 | 1.18 | 0.720 | 15 | 10.2 |

$$GF = 1.70 \times 10^{-4} \text{ m}^2\text{st} \qquad *j>1500 = P/m^2 \cdot s \cdot sr$$

$$t = 3.87 \times 10^6 \text{ s} \qquad j = 10^{-3} \text{ } P/m^2 \cdot s \cdot sr \cdot MeV/nuc$$

| Charges | >1500 MeV/nuc | 1200-1500 MeV/nuc | 1000-1200 MeV/nuc | 800-1000 MeV/nuc | 600-800 MeV/nuc | 500-600 MeV/nuc |
|---|---|---|---|---|---|---|
| **TABLE 3** | | | | | | |
| **Ratios to Oxygen Nuclei Intensities** | | | | | | |
| B[+] | 0.307 | 0.292 | 0.301 | 0.289 | 0.269 | |
| C | 1.043 | 1.048 | 1.024 | 1.047 | 1.072 | |
| O | j=3.46* 1.00 | j=2.67* 1.00 | j=3.72* 1.00 | j=5.31* 1.00 | j=7.35* 1.00 | j=9.56* 1.00 |
| Mg | 0.211 | 0.228 | 0.242 | 0.204 | | |
| Si | 0.179 | 0.159 | 0.174 | 0.173 | 0.158 | |
| Fe | 0.100 | 0.102 | 0.105 | 0.113 | 0.098 | |

\* These values are those in Table 2A
[+] Ratio for B = B/C

| TABLE 4 | | | | |
|---|---|---|---|---|
| **Calculation of Spectral Indices above 1.5 GeV/nuc** | | | | |
| Charge | Events >1.5 GeV/nuc | $J_0^* > 1.5$ GeV/nuc | $j_0^*$ at 1.5 GeV/nuc | 1.5 $(j_0/J_0)$+1 |
| B | 546 | 1.11 | 0.671 | 1.91$\pm$0.18 |
| C | 1812 | 3.61 | 2.28 | 1.95$\pm$0.11 |
| O | 1730 | 3.46 | 2.18 | 1.94$\pm$0.13 |
| Mg | 390 | 0.765 | 0.512 | 2.00$\pm$0.26 |
| Si | 336 | 0.647 | 0.350 | 1.81$\pm$0.28 |
| Fe | 180 | 0.340 | 0.215 | 1.95$\pm$0.40 |

$J_0^*$=Particles/$m^2 \cdot$sr$\cdot$s

$j_0^*$= Particles/$m^2 \cdot$sr$\cdot$s$\cdot$GeV/nuc

| TABLE 5 | | | | | | | |
|---|---|---|---|---|---|---|---|
| **He** | | | | | | | |
| **E/nuc** | **C2-4 CHAN** | **CNTS** | **C/MeV** | **INT** | **De-con** | **1/t x GF\*** | **+INTENSITY** |
| 1500 | 29.77 | | | | | | |
| | | 2956 | 9.85 | 1.05 | 0.66 ± 0.05 | 1/6.62 | 0.103 |
| 1200 | 31.10 | | | | | | |
| | | 2486 | 12.43 | 1.05 | 0.77 ± 0.04 | 1/6.62 | 0.152 |
| 1000 | 32.50 | | | | | | |
| | | 3020 | 15.10 | 1.05 | 0.89 ± 0.03 | 1/6.62 | 0.213 |
| 800 | 34.70 | | | | | | |
| | | 4010 | 20.05 | 1.05 | 0.95 ± 0.02 | 1/6.62 | 0.302 |
| 600 | 38.63 | | | | | | |
| | | 2746 | 27.46 | 1.05 | 0.98 ± 0.02 | 1/6.62 | 0.406 |
| 500 | 42.00 | | | | | | |
| | | 3235 | 32.35 | 1.05 | 1.00 | 1/6.62 | 0.513 |
| 400 | 46.90 | | | | | | |
| | | 4503 | 45.03 | 1.05 | 1.00 | 1/6.62 | 0.714 |
| 300 | 55.30 | | | | | | |
| | | 3577 | 59.5 | 1.05 | 1.00 | 1/6.62 | 0.944 |
| 240 | 64.16 | | | | | | |
| | | 2805 | 70.2 | 1.05 | 1.00 | 1/6.62 | 1.11 |
| 200 | 72.7 | | | | | | |
| | | 3316 | 82.9 | 1.05 | 1.00 | 1/6.62 | 1.315 |
| 160 | 86.8 | | | | | | |
| | | 1866 | 93.3 | 1.06 | 1.00 | 1/6.62 | 1.50 |
| 140 | 96.8 | | | | | | |
| | | 1096 | 54.8 | 1.06 | 1.00 | 1/3.31 GFF | 1.755 |
| 120 | 112.0 | | | | | | |
| | | 608 | 60.8 | 1.06 | 1.00 | 1/3.31 GFF | 1.94 |
| 110 | 122.6 | | | | | | |
| | | 632 | 63.2 | 1.06 | 1.00 | 1/3.31 GFF | 2.02 |
| 100 | 136.7 | | | | | | |
| | | 641 | 64.1 | 1.06 | 1.00 | 1/3.31 GFF | 2.055 |
| 90 | 157.0 | | | | | | |
| | | 660 | 66.0 | 1.06 | 1.00 | 1/3.31 GFF | 2.12 |
| 80 | 192.0 | | | | | | |

\* Livetime = 3.89 x $10^6$ sec          +Intensity - Particles/($m^2 \cdot sr \cdot$(MeV/nuc)) ± 5%

\* Geometry Factor = 1.7 x $10^{-4}$ $m^2 \cdot$ sr          CRITERIA:  B1 ≥ 15, C1 ≥ 14, 1.48 ≤ B1/C1 ≤ 1.80

GFF = Forward Geometry Factor = 0.85 x $10^{-4}$ $m^2 \cdot$ sr

# REFERENCES

Adriani, O., Barbarino, G.C., Bazilevskaya, G.A., et al., 2013, ApJ, <u>770</u>, 2

AMS DAYS at CERN, 2015

Cowsik, R. and Wilson, L.W., 1975, Proc. 14[th] ICRC (Munich), <u>2</u>, 659

Cummings, A.C., Stone, E.C., Heikkila, B.C., et al., 2016, Ap.J., <u>831</u>, 21

Engelmann, J.J., Ferrando, P., Soutoul, A., et al., 1990, A&A, <u>233</u>, 96

Garcia, Munoz, M., Simpson, J.A., Guzik, T.G., et al., 1987, ApJS, <u>64</u>, 269

Lave, K.A., Wiedenbeck, M.E., Binns, W.R., et al., 2013, Ap.J., <u>770</u>, 117

Schlickeiser, R. and Lerche, I., 1985, ICRC, <u>3</u>, 54S

Stone, E.C., Vogt, R.E. and McDonald, F.B., et al., 1977, Space Sci. Rev., <u>21</u>, 359

Usoskin, I.G., Bazilevskaya, G.A. and Kovaltson, G.A., 2011, J. Geophys. Res., <u>116</u>, A02104

Webber, W.R., McDonald, F.B. and Lukasiak, A., 2003, Ap.J., <u>599</u>, 582-595

Webber, W.R. and Higbie, P.R., 2015, http://arXiv.org/abs/1503.05891

Webber, W.R., 2017, http://arXiv.org/abs/1711.11584

Webber, W.R. and Villa, T.L., 2016, http://arXiv.org/abs/1606.03031

Webber, W.R. and Villa, T.L., 2017, http://arXiv.org/abs/1711.08015

**FIGURE CAPTIONS**

**Figure 1:**  Outline drawing of identical high energy bi-directional telescopes on V1 and V2.

**Figure 2:**  Schematic drawing of penetrating particle pulse heights of different Z that lie along the diagonal in a cross plot matrix of C2-4 vs. C1 events.  The location of the various energy particles for each charge are calculated using the energy loss program.

**Figure 3:**  Pulse height distribution of B-O nuclei measured at V1 beyond the HP from 2012.9 to 2017.0.

**Figure 4:** Spectrum of B nuclei measured by Voyager 1 between a few MeV/nuc to ~3 GeV/nuc. Solid red circles above 0.5 GeV/nuc are data reported in this paper.  Open red circles are from W, M and L, 2003.  HEAO-3 measurements (Engelmann, et al., 1990) are shown as solid blue circles.  PAMELA (Adriani, et al., 2013) as solid green circles.  The V1 measurements at lower energies (Cummings, et al., 2016) are shown as solid red circles and squares.  The solid black lines are truncated LBM calculations with $P_0 = 0.562$ GV and $\lambda_2 = 0.12$ as described in the text.  The dashed blue line is the GALPROP diffusive re-acceleration model from Cummings, et al., 2016.

**Figure 5:**  The same as Figure 4 but for C nuclei. Dashed black lines are for solar modulation = 400 MV and 660 MV

**Figure 6:**  The same as Figure 4 but for O nuclei along with HEAO-3 data and solar modulation = 660 MV

**Figure 7:**  That same as Figure 6 but for Mg nuclei.

**Figure 8:**  The same as Figure 7 but for Si nuclei.

**Figure 9:**  The same as Figure 7 but for Fe nuclei.

**Figure 10:**  Distribution of events in C2-4 along the diagonal for He nuclei.  Also shown along the bottom is the energy loss in terms of the He energy in MeV/nuc.

**Figure 11:**  Distribution of large pulse height events in C2-4 for He nuclei showing the separation of forward (F) and backward (B) events at lower energies.  Shaded region is energies where the F/B fraction is uncertain.

**Figure 12:**  LIS He spectrum from 80 to 1500 MeV/nuc from this study and also that published for lower energies (Cummings, et al., 2016).  The calculated He spectrum for a source rigidity spectrum with exponent = -2.24 and an interstellar diffusion coefficient ~rigidity$^{0.45}$ above $P_0$=0.5 GV is shown for a LBM with a truncation parameter = 0.04 (see text).

**Figure 13:**  Observed relative intensities of He, C, O, Mg, Si and Fe nuclei between 100 and 1000 MeV/nuc.  These intensities are normalized to the values of j at 1000 MeV/nuc for Carbon nuclei (see Tables 2 and 5).  The figure includes lower energy intensities from Cummings, et al., 2016.

**Figure 14:**  Ratios of intensities at 100 MeV/nuc to those at 1 GeV/nuc for He, C, O, Mg, Si and Fe nuclei.  The calculated ratios for the same truncated LBM propagation parameters as used for Figures 4-9 and Figure 12 are shown as horizontal solid black lines for source spectral indices between $P^{-2.24}$ and $P^{-2.28}$.  The red points directly above these lines are the measurements and errors.

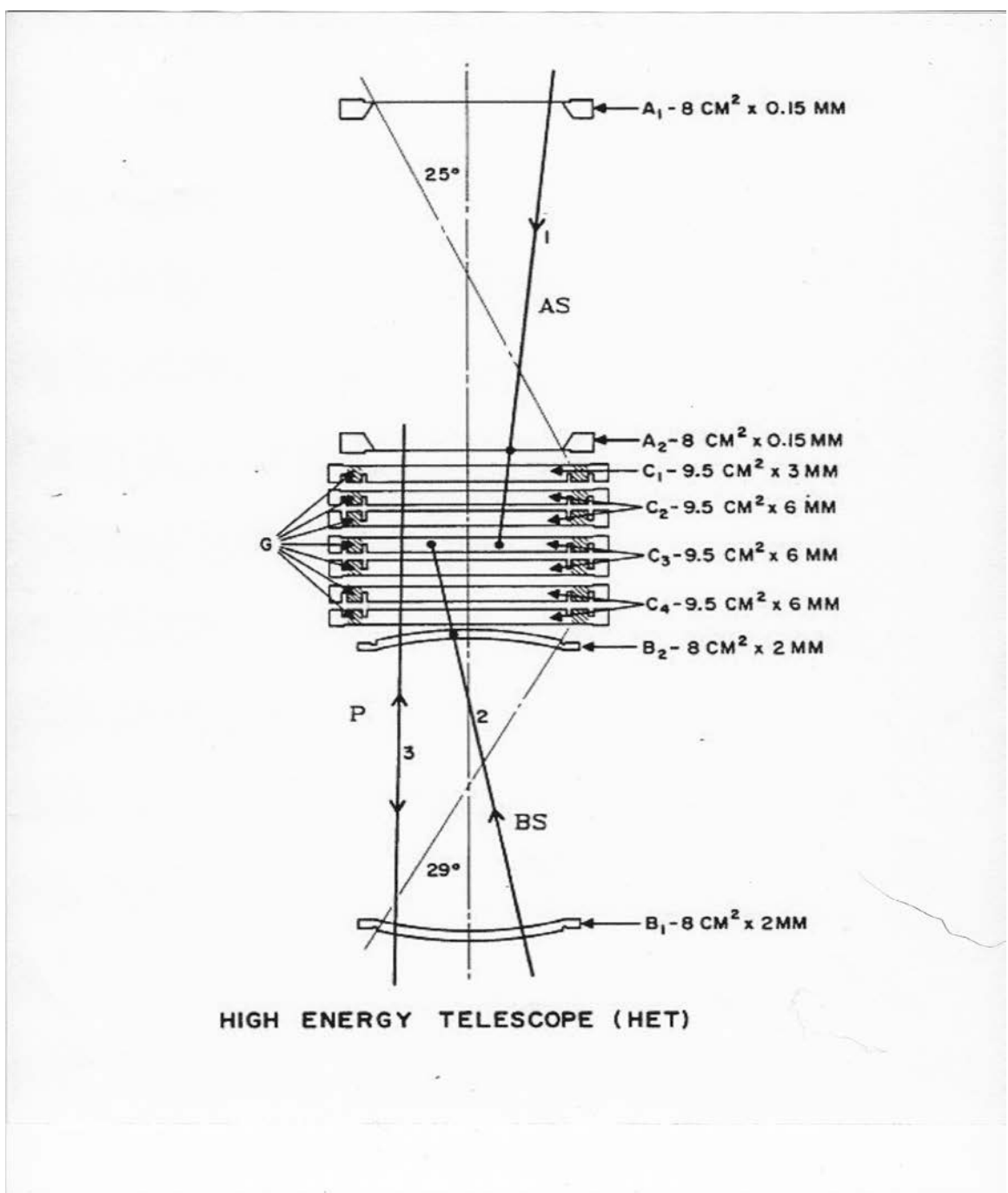

**FIGURE 1**

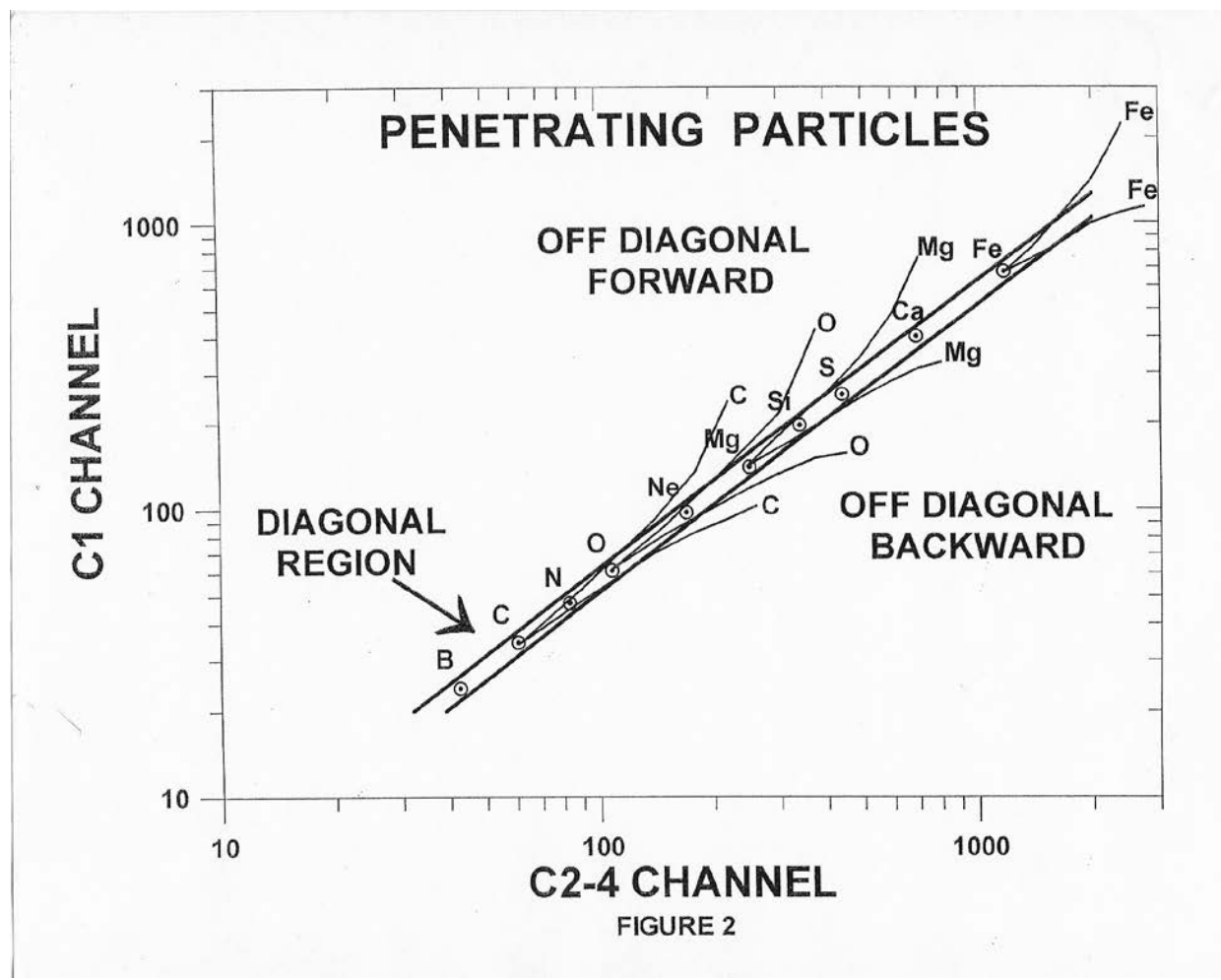

FIGURE 2

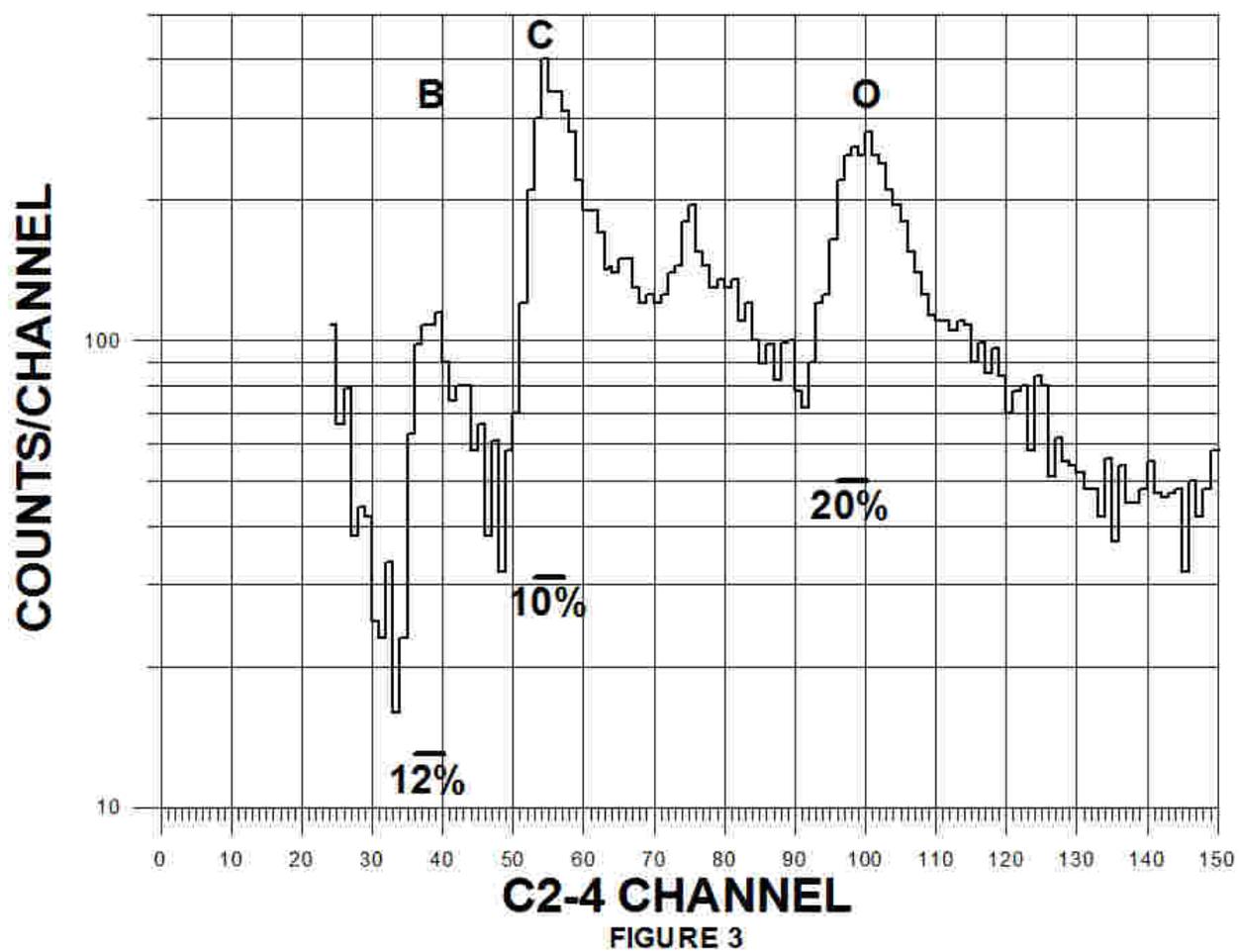

FIGURE 3

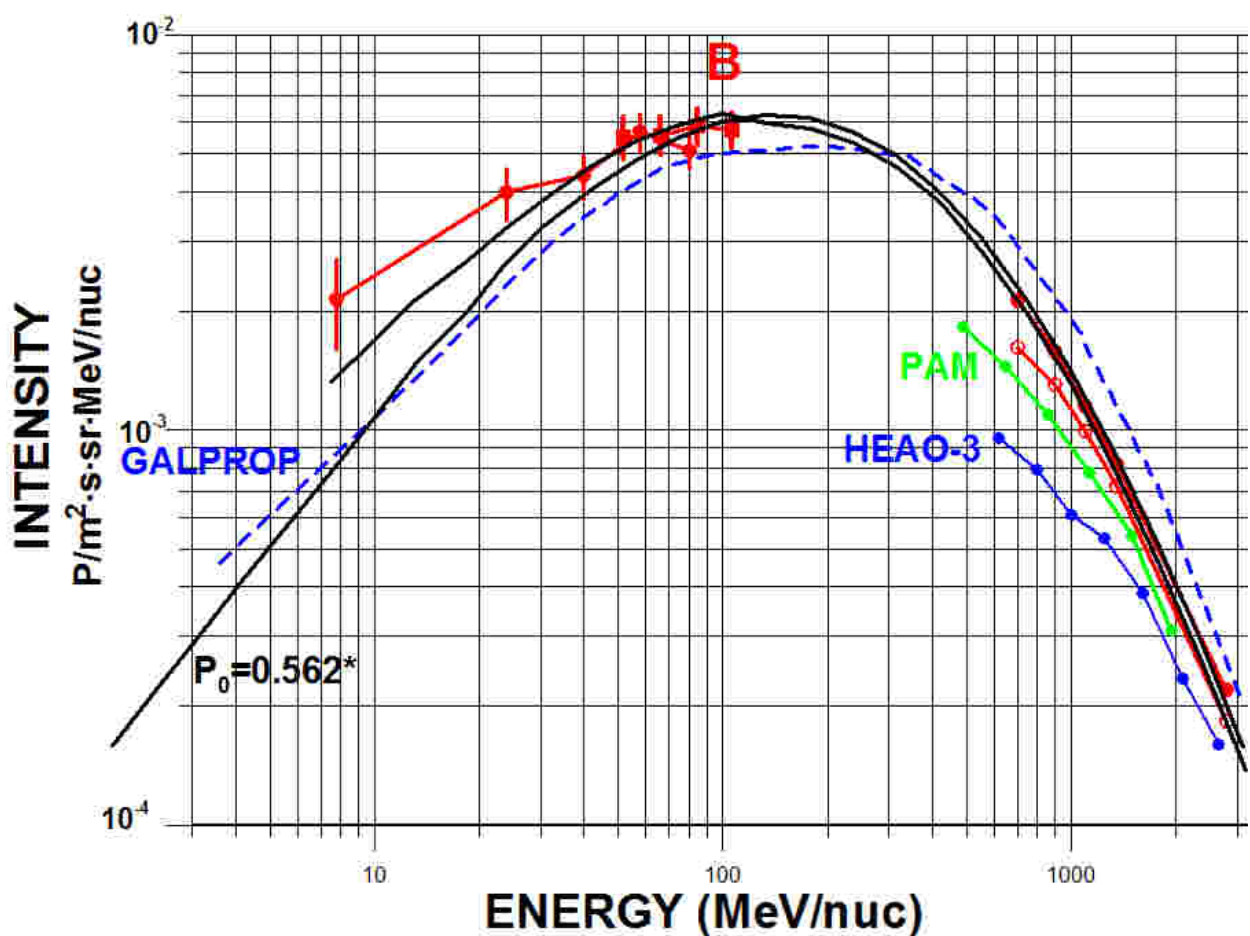

FIGURE 4

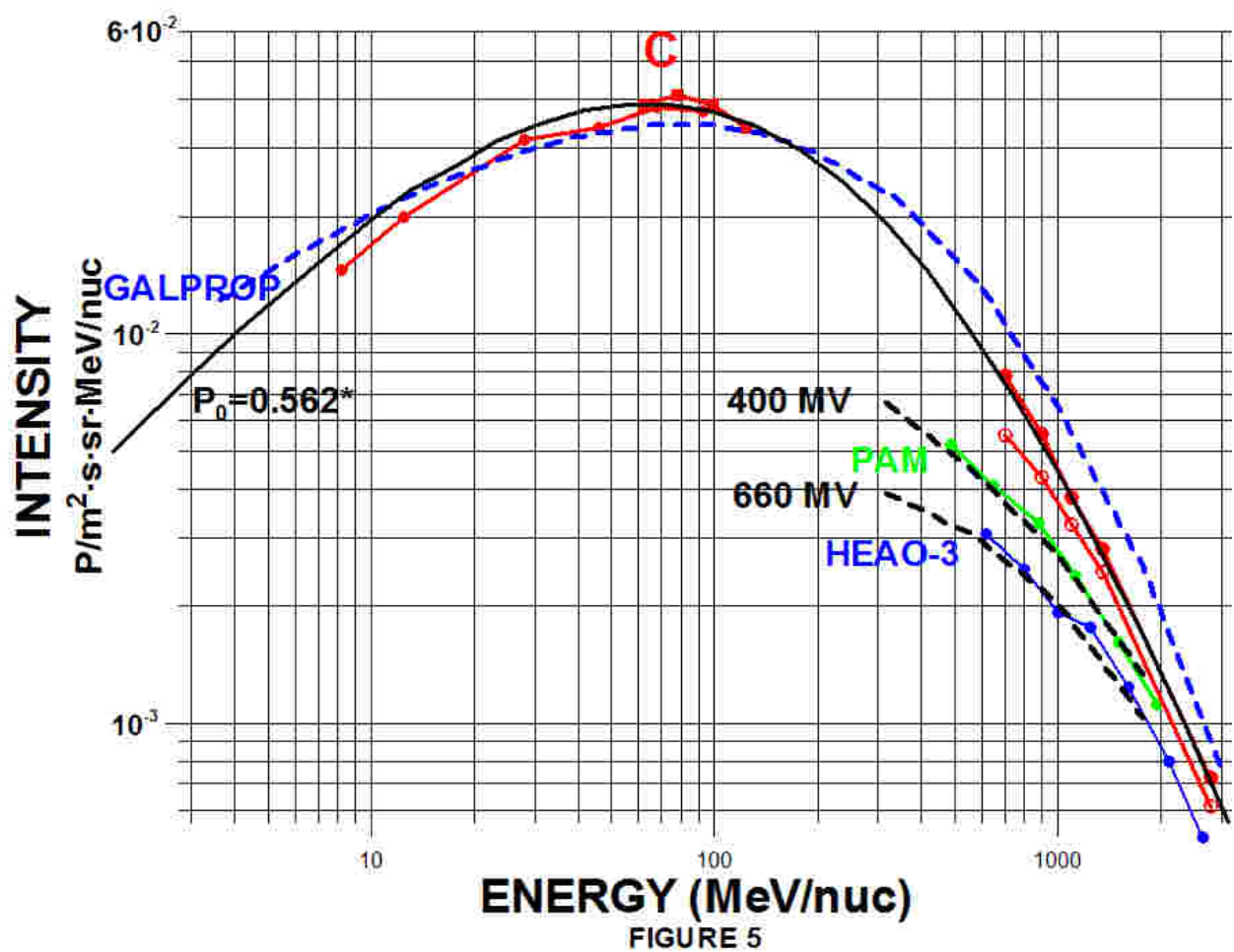

**FIGURE 5**

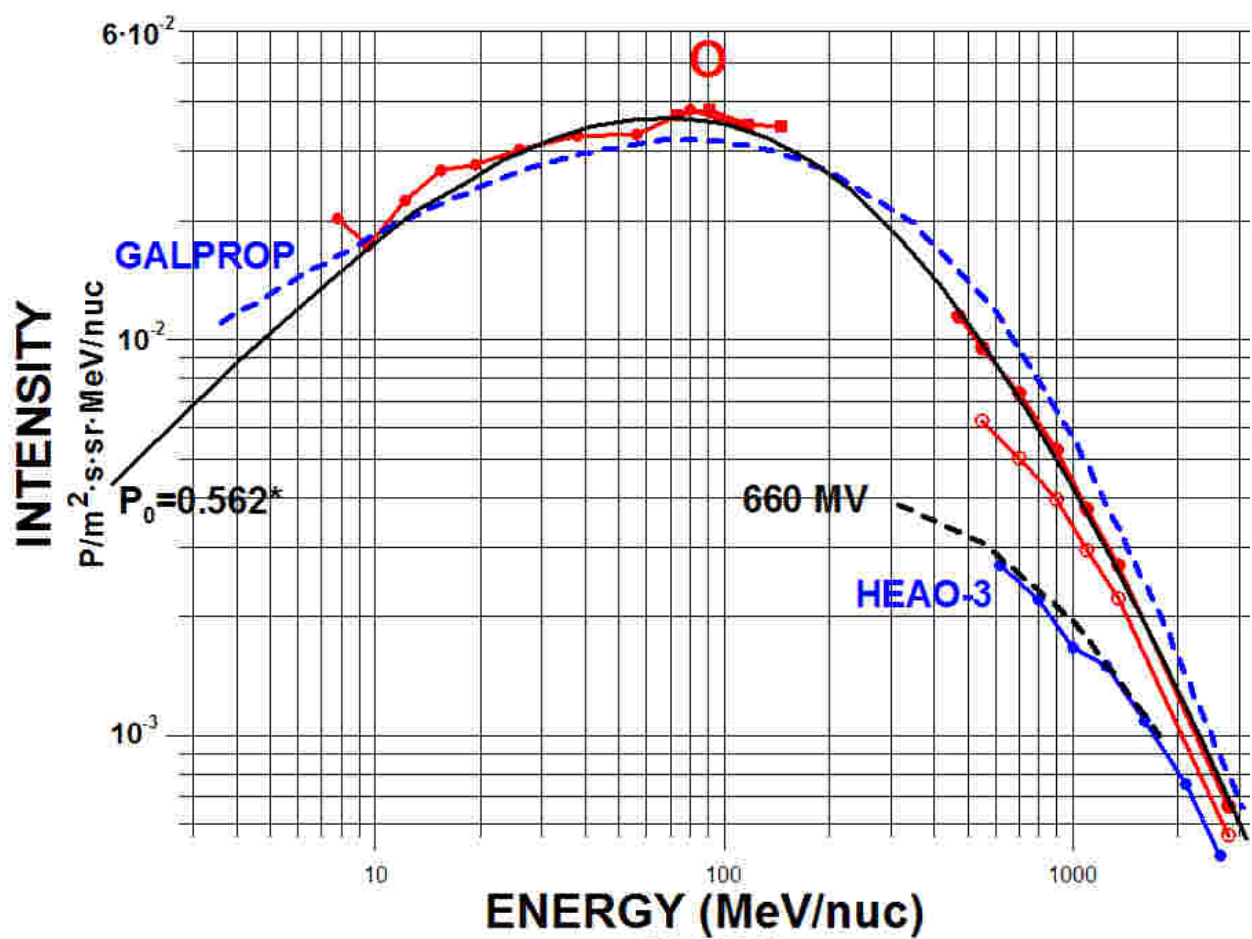

FIGURE 6

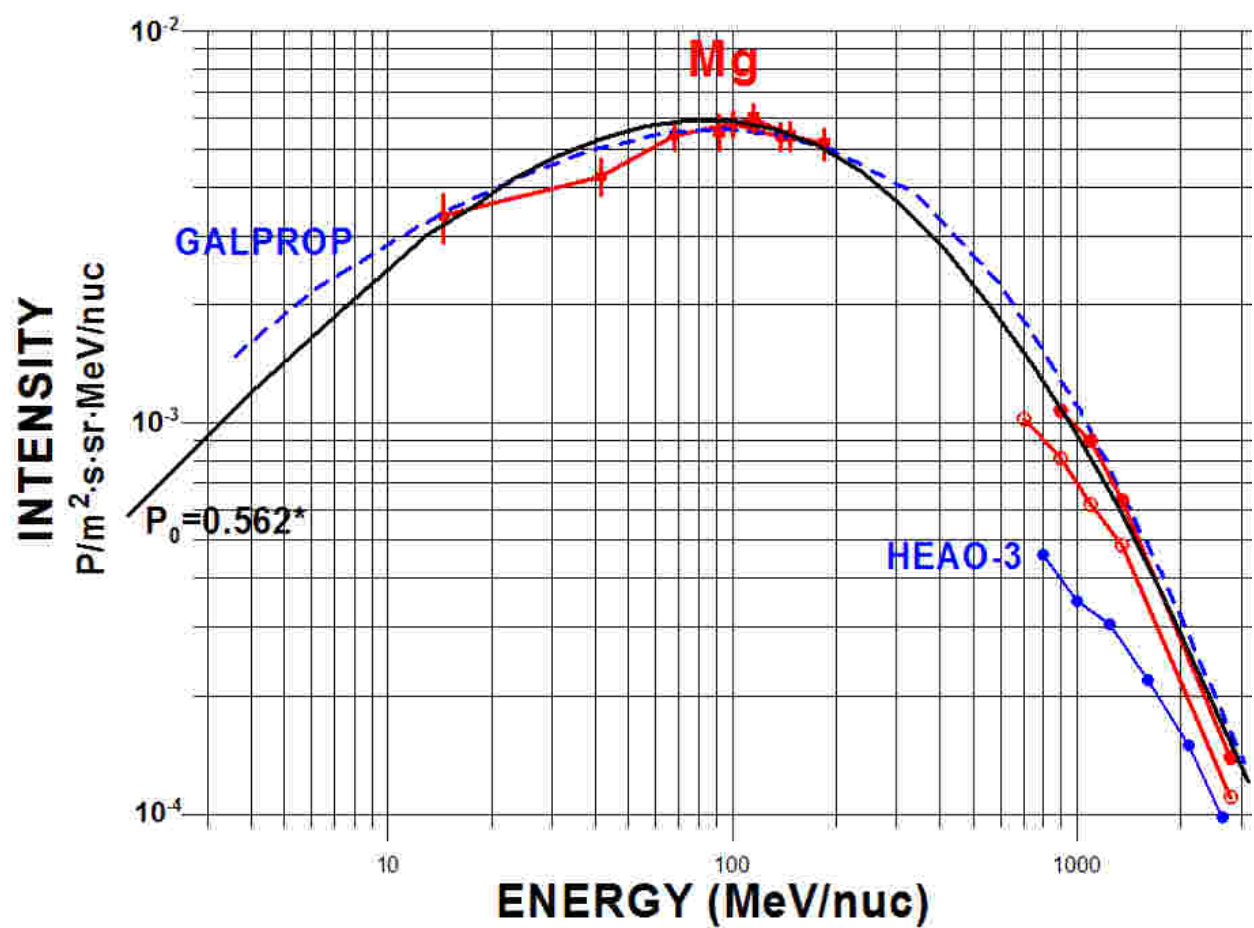

**FIGURE 7**

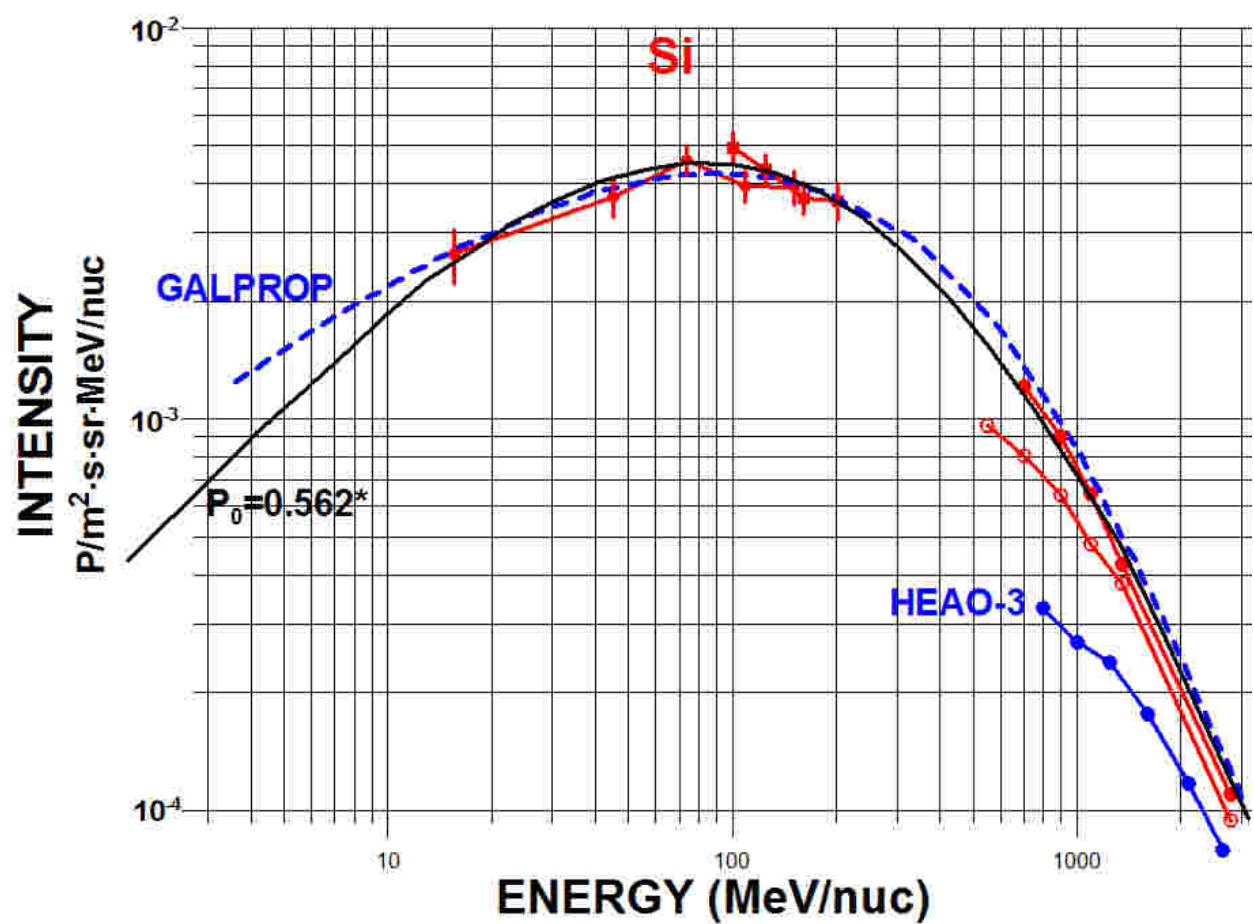

**FIGURE 8**

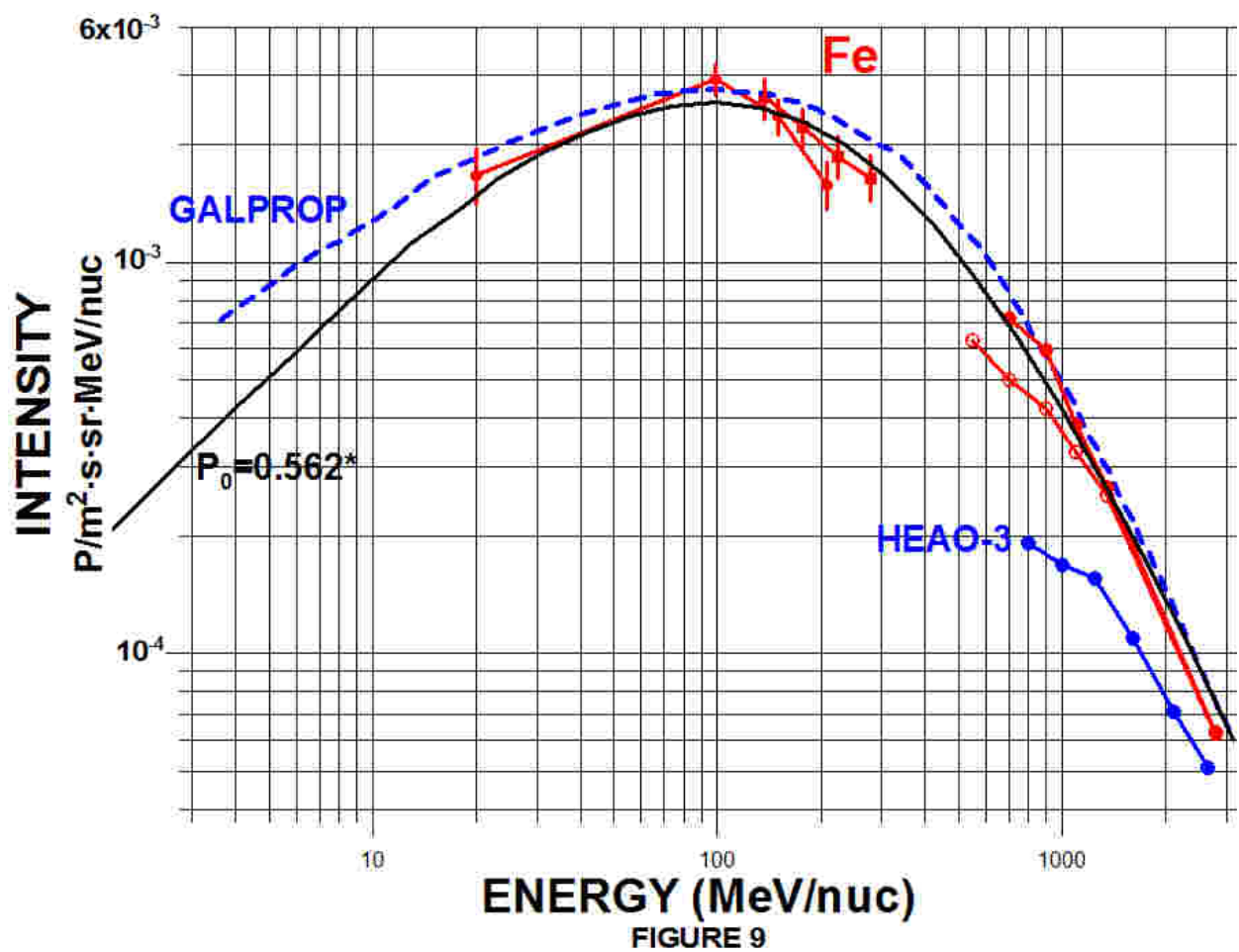

**FIGURE 9**

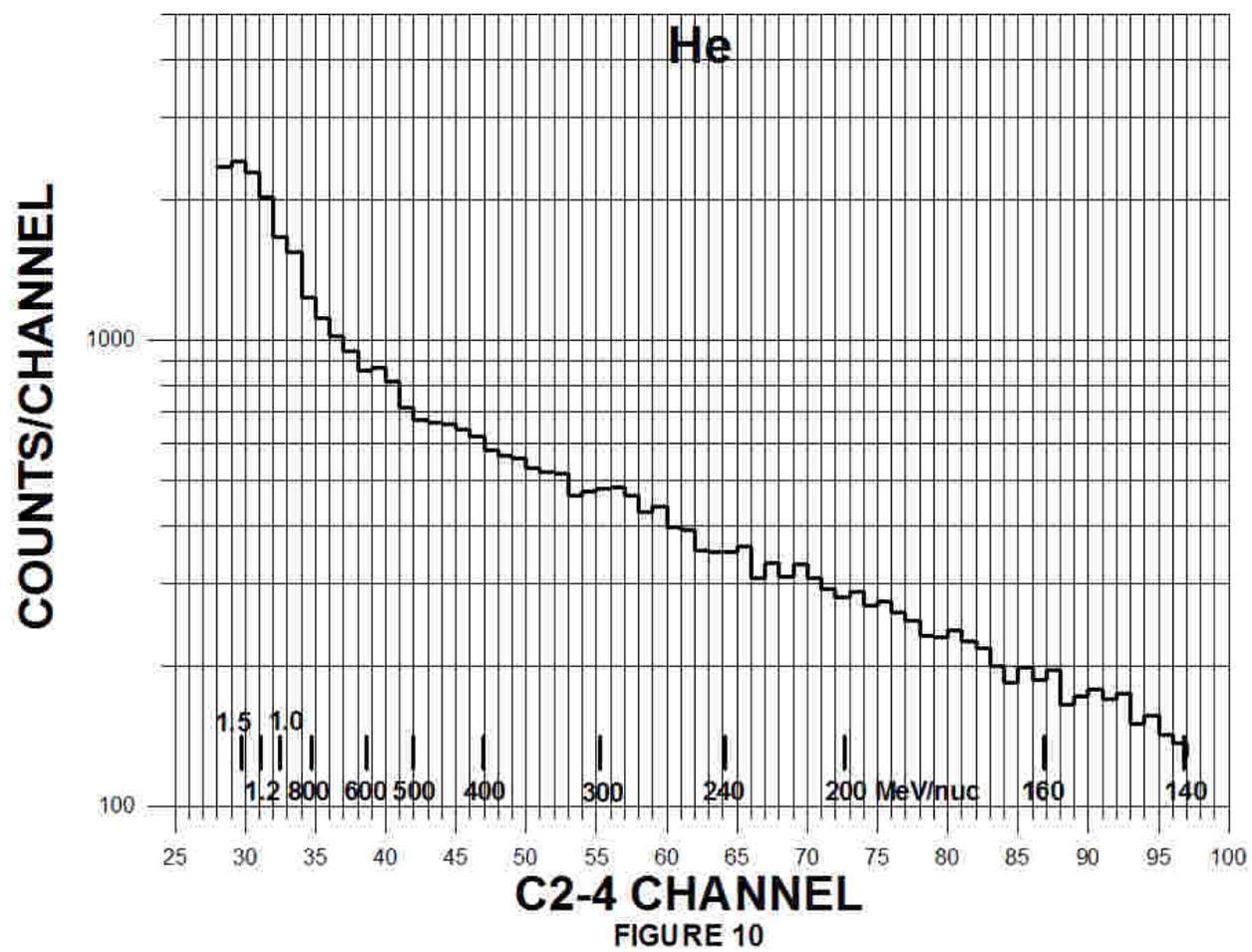

FIGURE 10

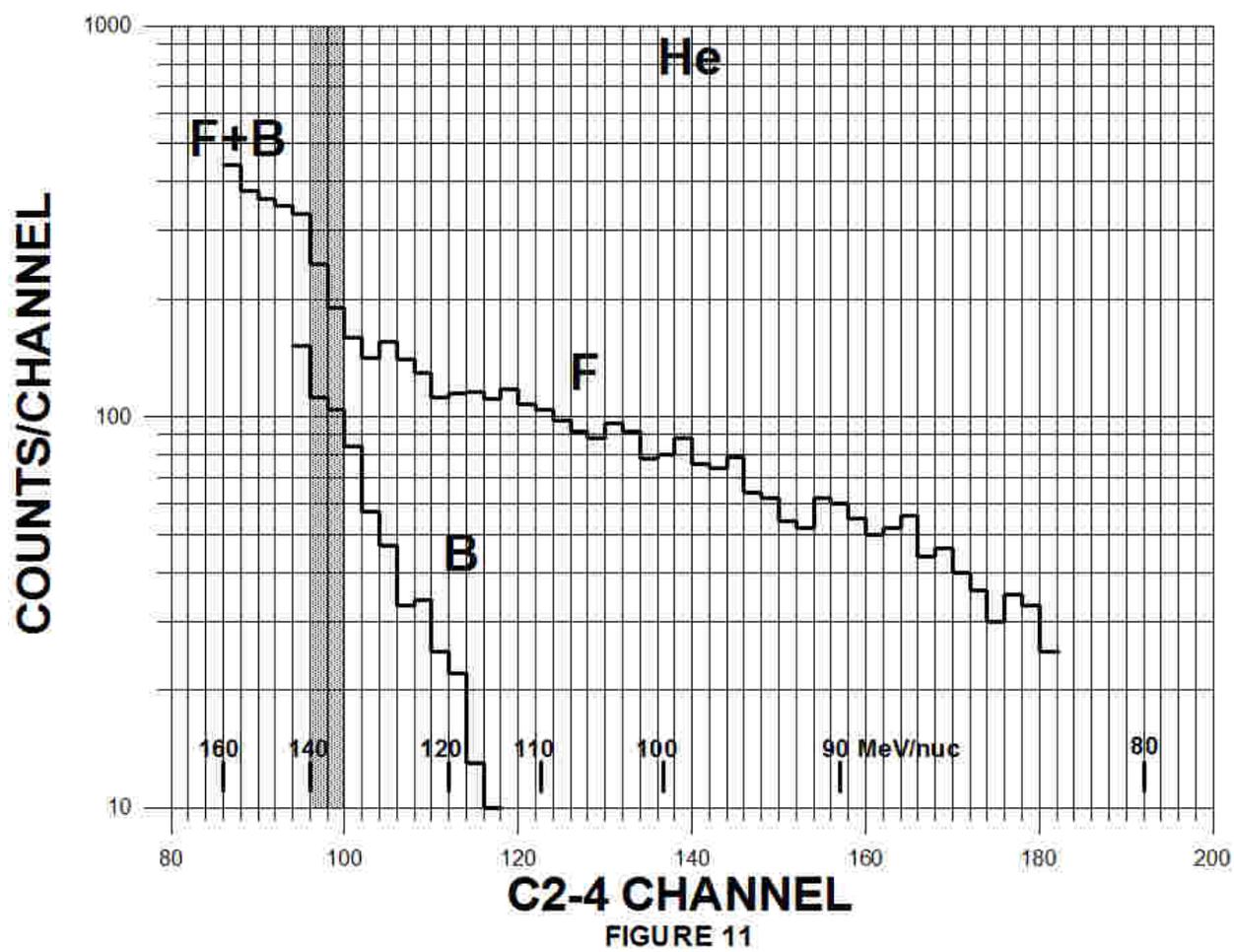

FIGURE 11

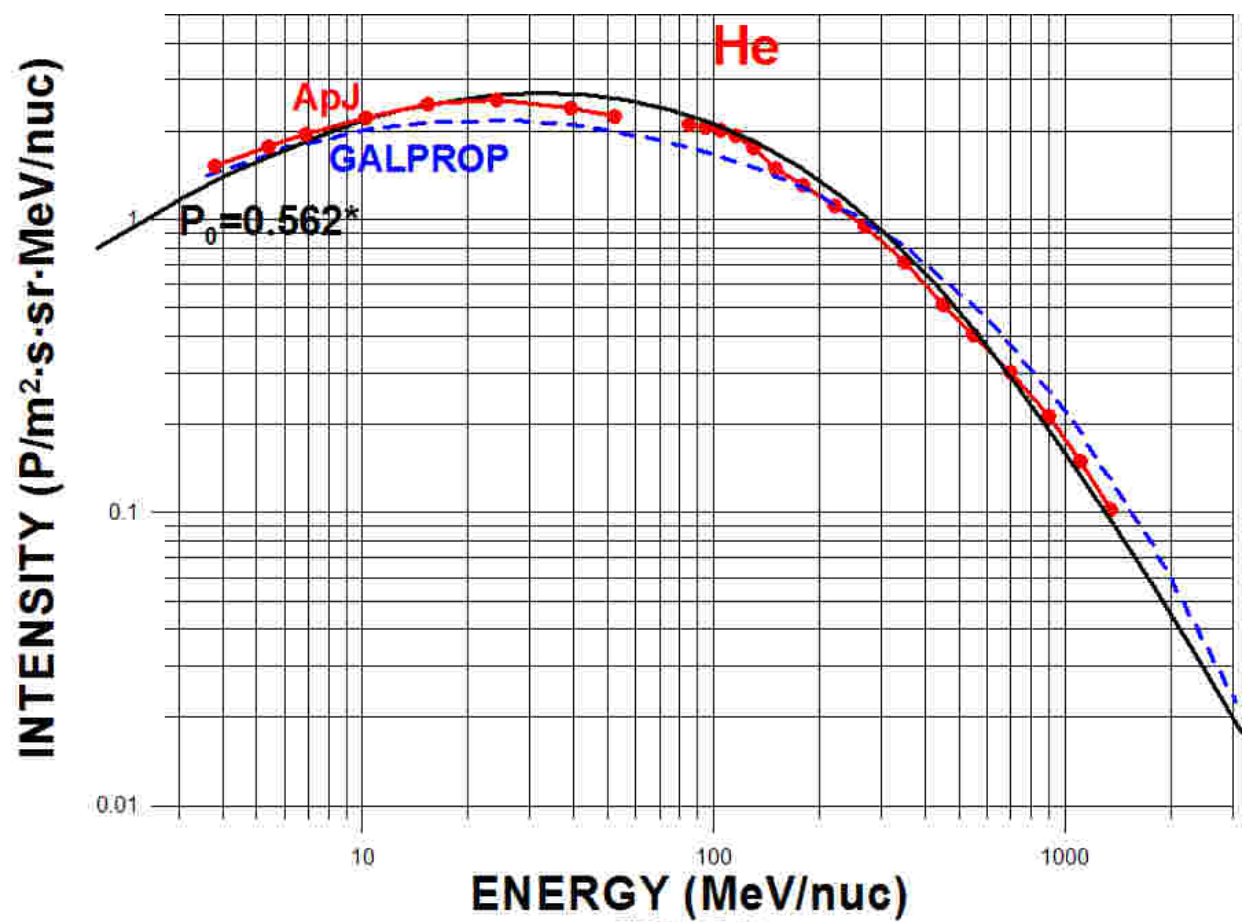

FIGURE 12

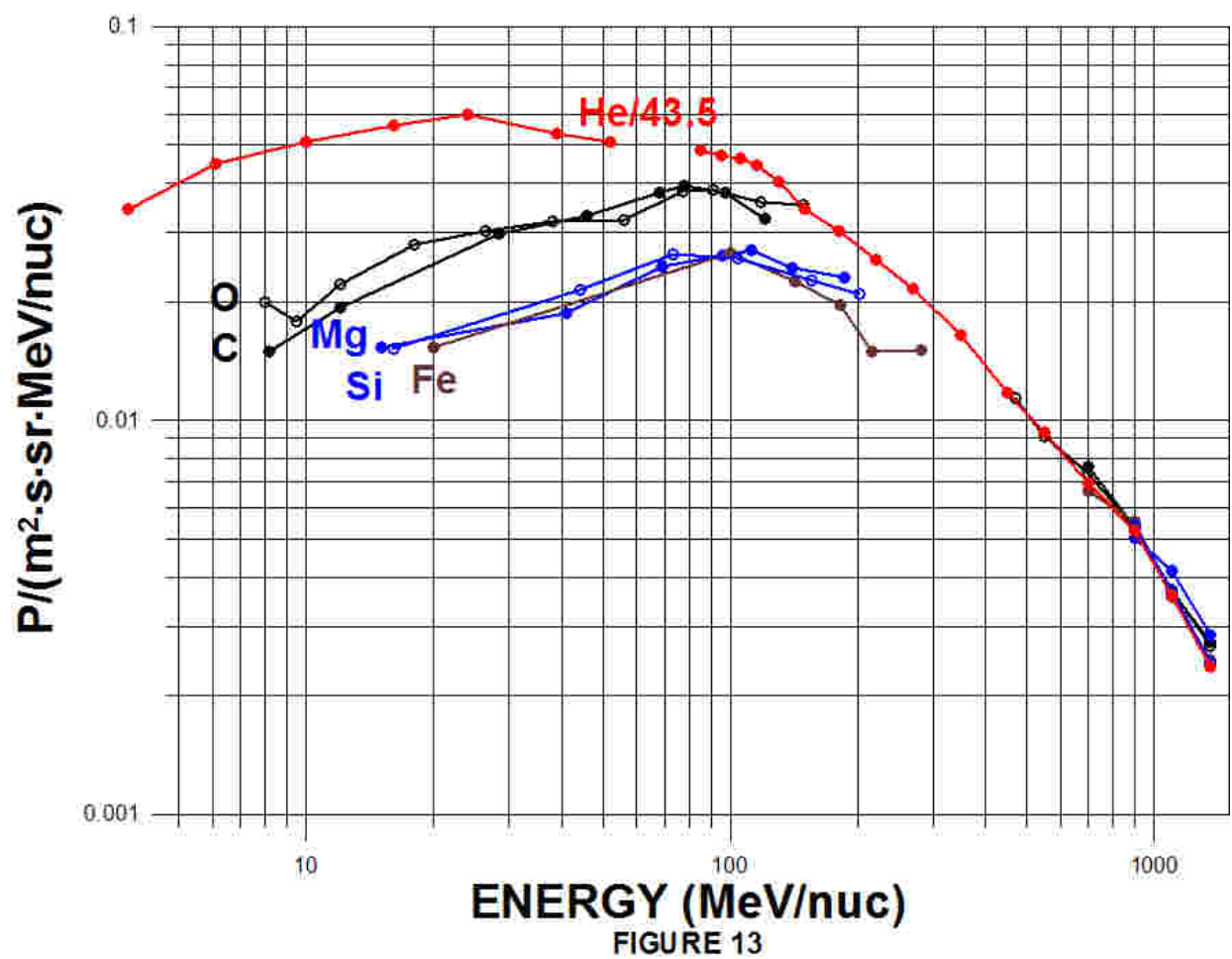

**FIGURE 13**

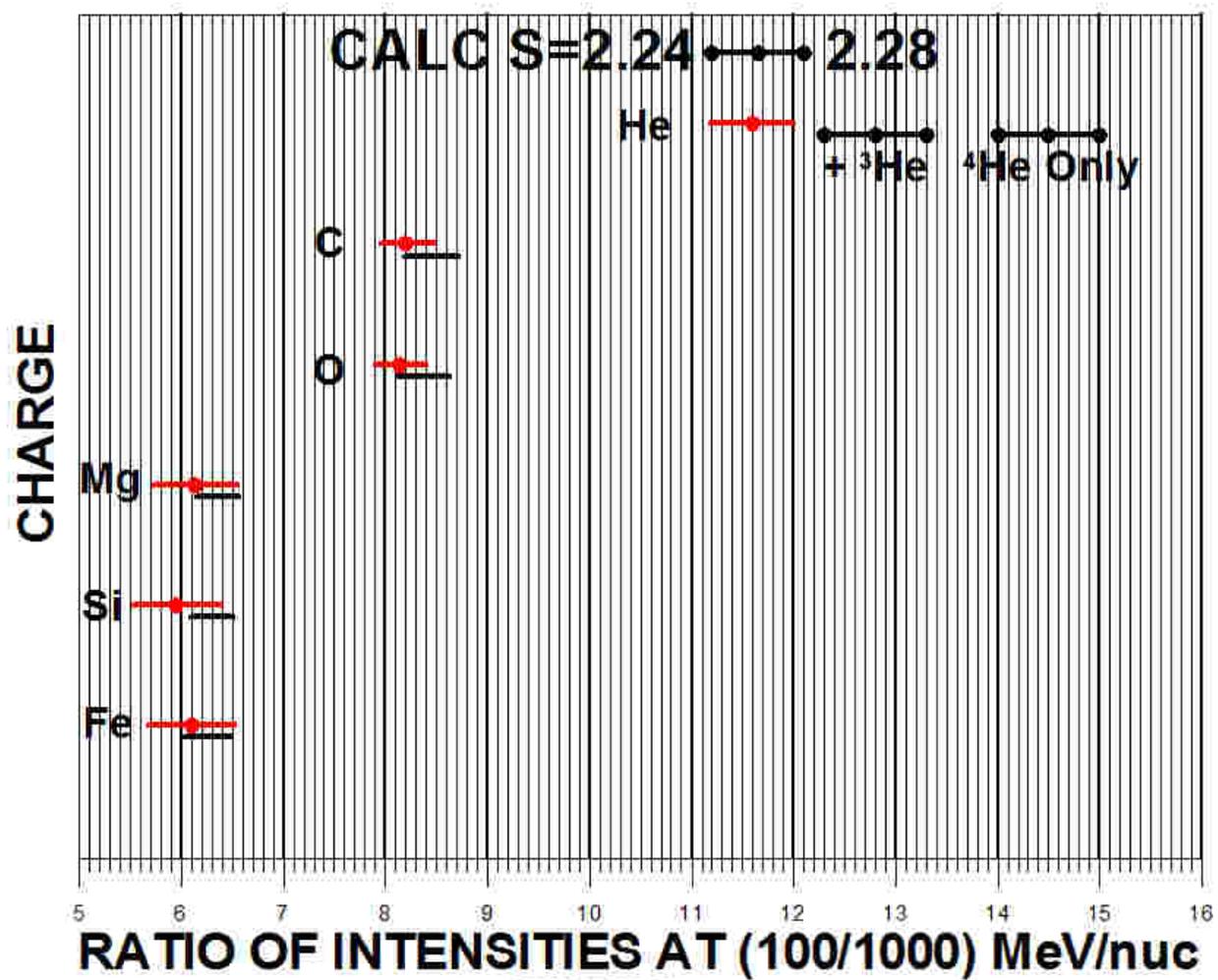

FIGURE 14